\documentclass[%
 reprint,
superscriptaddress,
 amsmath,amssymb,
 aps,
]{revtex4-1}

\usepackage{graphicx}% Include figure files
\usepackage{dcolumn}% Align table columns on decimal point
\usepackage{bm}% bold math
\usepackage{hyperref}
\usepackage{siunitx}
\usepackage{xcolor}
\usepackage{systeme}
\newcommand{\erfc}{\text{erfc}}

\begin{document}

%\preprint{APS/123-QED}

\title{
Disjoining pressure oscillations causing height discretization in graphene nanobubbles
}% Force line breaks with \\
%\thanks{}
\author{Timur Aslyamov}
\email{t.aslyamov@skoltech.ru}
\affiliation{Center for Design, Manufacturing and Materials,
  Skolkovo Institute of Science and Technology,
  Bolshoy Boulevard 30, bld. 1, Moscow, Russia 121205}
\author{Ekaterina Khestanova}
\affiliation{ITMO University, Saint Petersburg, 197101, Russia}
\author{Maria Korneva}
\author{Evgeny Iakovlev}
\author{Petr Zhilyaev}
\author{Iskander Akhatov}
\affiliation{Center for Design, Manufacturing and Materials,
  Skolkovo Institute of Science and Technology,
  Bolshoy Boulevard 30, bld. 1, Moscow, Russia 121205}
\date{\today}% It is always \today, today,

\begin{abstract}
Recent experiments and computer simulations observe various geometrical formations of nanobubbles in van der Waals heterostructures. Among the well studied dome and tent geometries, there is yet least understood pancake graphene nanobubbles (GNB). This more exotic form exhibits discrete values of vertical sizes around just a few diameters of the molecules trapped inside the GNBs. We develop a model based on the membrane theory and confined fluids thermodynamics. Our approach describes  the equilibrium properties of such flat GNBs. We show that discrete pancake geometry is the result of disjoining pressure induced by the trapped fluid inside GNB. The calculated total energy defines a discrete series of the metastable states with the pancake heights, which are multiple to molecular diameter. We observe that the value and the distribution of the total energy minima crucially depend on the temperature. The energy barriers between metastable states decrease as the temperature becomes larger. 
Also, we demonstrate that the pancake forms are favorable in the cases of sufficiently low membrane-substrate adhesion energy and the small number of trapped molecules. These properties are in agreement with the published simulations and experiments. The numerical comparison of our result with molecular dynamics results additionally shows the adequacy of the proposed model.
\end{abstract}

\maketitle

\section{introduction}
Experimental study \cite{khestanova2016universal} demonstrated that the molecules trapped between two-dimensional crystal and the substrate form the graphene nanobubbles (GNBs). The measured GNBs profiles  exhibit a smooth dome form and correspond to the universal scaling law derived from the elastic membrane theory \cite{khestanova2016universal}. Beside smooth geometry, two dimensional materials may form sharp tents, which also exhibit universal shape characteristics \cite{dai2018interface}. Therefore in the dependence on the geometry these nanostructures can be described by the spherical membrane model \cite{yue2012analytical} or tent geometry modifications \cite{dai2018interface}. However, established membrane and GNB models do not cover pancake (flat island) geometry, which are recently observed in computer simulations and atomic field microscopy experiments \cite{iakovlev2017atomistic, khestanova2018van}. In this case the trapped molecules are packed in layers and such GNB profile exhibit a notable flat region Fig.\ref{fig:profile_displacement}. The observed pancakes have extremely small vertical size, which is multiple of the trapped molecules diameter. Thus, instead of the universal scaling the the pancakes show the discrete geometry induced by the properties of the trapped molecules. 

Molecular dynamic (MD) simulation of small number of argon molecules ($N=2500$) between two graphene sheets demonstrated the existence of the stable pancakes underneath graphene surface \cite{iakovlev2017atomistic}. These calculations observed at least two equilibrium states: the first, more stable, GNB configuration corresponds to the pancake with the vertical size around two argon diameters; the second one is the intermediate state from the pancake to the dome form, where the flat region is notable but the aspect ratio fits the observed early round GNBs \cite{khestanova2016universal}. The simulation shows the metastability of the pancake equilibrium states with discrete vertical size. 
Other MD simulations \cite{sanchez2018mechanics} showed that water-filled blisters exhibit both dome and pancake forms in the dependence on the adhesion energy and the number of trapped water molecules. The flat shape was observed for blisters with the height lower than three molecular diameters in the cases of relatively low adhesion energy and the number of trapped molecules \cite{sanchez2018mechanics}. 

Unlike dome and tent forms, the pancakes are observed at nanoscopic scales only, when the vertical size is several molecular diameters. Such size limitations point out the trapped molecules significantly influence the GNB characteristics. As a result the pancakes vertical size holds the discrete values, which are multiple for the diameter of the trapped molecules. Therefore, to extend GNB theory to the pancake geometry, both modifications of elastic membrane theory and analysis of the confined fluid properties are necessary. In this work, we develop a theory of pancakes consisting of two sheets of two-dimensional crystals and trapped simple fluid at various temperatures. Our approach is based on both the model of elastic deformations of the flat-shape profile and confined fluid equation of state accounting for the impact of the disjoining pressure. The nanoscale pressure of confined fluid exhibits oscillations with the period equals to the molecular diameter and tends to bulk pressure as confinement size increases \cite{israelachvili2015intermolecular}. This deviation from the bulk pressure is limited by only a few nanometers length scale, but may achieve the values around GPa \cite{long2011pressure}. Also such short range phenomena plays important role in the description of the equilibrium of liquid films and gas nanobubbles on the surface \cite{svetovoy2016effect,kim1999disjoining}. Therefore, for the pancakes of nanoscale size the effects of confined fluids influence crucially their equilibrium. To obtain numerical results, we use Classical Density Functional Theory (c-DFT), which can provide the disjoining pressure \cite{gregoire2018estimation}, the confined fluid density \cite{wu2006density} and the corresponding  energy simultaneously \cite{aslyamov2020model}. 
 
As example we consider argon molecules trapped between two graphene sheets at temperature range from \si{250}{K} till \si{450}{K}. Using our approach, we reveal the conditions of the equilibrium pancakes. Also, we investigate the influence of such parameters as temperature, membrane-substrate adhesion energy, and the number of trapped molecules. To verify the developed approach, we quantitatively compare our results with MD simulations. Developed model provides the following properties of the pancakes, in agreement with previous simulations and experiments:

\begin{itemize}
    \item The disjoining pressure oscillations induce the GNB states with the pancake profiles and discrete vertical sizes.
    \item The pancakes are metastable states corresponding to the local minima of the total energy. The depth of the energy minimum decreases as the height becomes larger. There are no pancakes with more than three layers of the trapped molecules inside. 
    \item The temperature and membrane-substrate adhesion energy extremely influence the geometrical characteristics of the GNB. The increase of these parameters collapses the pancakes, especially in the case of three layers GNB. 
    \item Heating to a sufficiently high temperature and then cooling to the initial temperature induces the transition of the pancake state to the dome shape.   
\end{itemize}

\section{model}
The general expression of the GNB total energy is well known from the membrane theory \cite{khestanova2016universal} and contains the following three terms: 
\begin{equation}
\label{eq:total_energy}
E_\text{tot}=E_\text{el}+E_\text{cf}+E_\text{adh}
\end{equation}
where $E_\text{el}$ is the elatic contribution, $E_\text{cf}$ is the confined fluid energy, $E_\text{adh}$ is the adhesion energy. Ordinarily term $E_\text{cf}$ is described by bulk equation of state that is not appropriate at nanoscale due to inhomogeneous fluid distribution in normal to the surface direction. In our study $E_\text{cf}$ contains the energy of fluid trapped inside GNB and the contribution from the interaction between fluid molecules with both substrate and membrane. Since fluid-solid adhesion energy is already accounted by $E_{cf}$, the term $E_{adh}$ corresponds to membrane-substrate interaction only. Thus the energy of the adhesion is defined as $E_\text{adh}=\pi L^2 \gamma$, where $L$ is the footpint radius, $\gamma$ is the specific adhesion energy between the graphene sheet and the substrate. 

To describe the elastic properties of the graphene enclosed pancakes, we modify the approach, which is successfully used for the smooth spherical \cite{yue2012analytical} and sharp tent geometries \cite{dai2018interface}. Similarly, with the work \cite{dai2018interface}, we describe the pancake profile by model explicit function $h(r)$, where $r$ is the lateral radial coordinate. 

\begin{figure}
    \centering
    \includegraphics[width=0.45\textwidth]{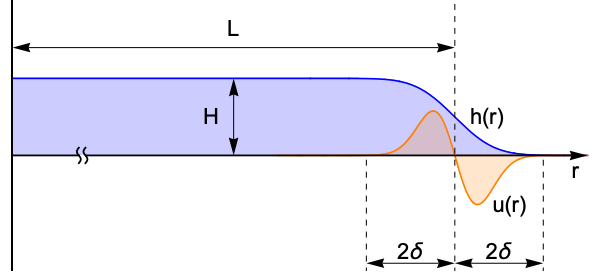}
    \caption{\label{fig:profile_displacement}The sketch of the characteristic pancake profile (blue line) defined by expression (\ref{eq:profile_expression}) and calculated corresponding in-plane distribution line from formula (\ref{eq:displacements}).}
\end{figure}

We define the profile $h(r)$ as the following expression:
\begin{equation}
\label{eq:profile_expression}
h(r)=\frac{H}{2}\erfc\left(\frac{r-L}{\delta}\right)
\end{equation}
where $H$ is the height of the GNB, $L$ is the footprint radius, $\delta$ is the characteristic length of the transition region between the membrane and the substrate (see Figure~\ref{fig:profile_displacement}).  

In accordance with the membrane theory the in-plane displacements corresponding to the certain profile can be defined from the in-plane momentum conservation law in polar coordinates \cite{wang2013numerical}:
\begin{equation}
\label{eq:momentum_conservation}
\frac{d^2 u}{d r^2}+\frac{1}{r}\frac{d u}{d r}-\frac{ u}{r^2}=-\frac{1-\nu}{2r}\left(\frac{d h}{d r}\right)^2-\frac{d h}{d r}\frac{d^2 h}{d r^2}
\end{equation}
where $\nu$ is the Poisson's ratio. As one can see the terms on the right hand of equation \eqref{eq:momentum_conservation} have a common exponential factor $e^{-2\left(\frac{r-L}{\delta}\right)^2}$, which provides significant value of solution $u(r)$ in the vicinity of $r=L$ only. Considering the singularity of equation \eqref{eq:momentum_conservation} at $r\to0$ and the expansions at $r\to L$ we  derive the following approximated in-plane displacement (the detailed calculations see S1):
\begin{equation}
\label{eq:displacements}
u(r)\simeq \frac{H^2r(L-r)}{4\pi\delta^2 L} \exp\left[-2\left(\frac{r-L}{\delta}\right)^2\right]
\end{equation}
The relation to the profile geometry (\ref{eq:profile_expression}) is shown in Figure~\ref{fig:profile_displacement}. 
To describe the profiles of real pancakes \cite{iakovlev2017atomistic} in terms of explicit function \eqref{eq:profile_expression} the length $\delta$ should be the function of the GNB size. More precisely, the length $\delta$ increases as the height $H$ becomes larger. For this reason, we use the following general power dependence $\delta=\delta_0 \left(\frac{H}{d}\right)^k$, where $\delta_0$ and $k$ are the parameters defining the curved region, $d$ is the diameter of the trapped molecules, which is used here to correct the length dimension.

The information about the profile and in-plane displacement is complete enough to calculate the elastic energy term as the following \cite{wang2013numerical}:
\begin{equation}
\label{eq:elstic_energy}
E_\text{el}=2\pi\int\frac{Y}{2(1-\nu^2)}\left(\epsilon_r^2+2\nu\epsilon_r\epsilon_\theta+\epsilon_\theta^2\right)rdr
\end{equation}
where $Y$ is two-dimensional Young modulus, $\epsilon_r$ and $\epsilon_\theta$ are the radial and circumferential strain components, which can be found using expressions (\ref{eq:profile_expression},\ref{eq:displacements}) and the following definitions \cite{yue2012analytical}:
\begin{equation}
\label{eq:strain_tensor}
\epsilon_r=\frac{d u}{d r}+\frac{1}{2}\left(\frac{d h}{d r}\right)^2, \quad \epsilon_\theta=\frac{u}{r}
\end{equation}
%Resulted elastic energy $E_\text{el}(H)$ is a function of the maximal height $H$ only. 

The molecules trapped inside GNB with nanoscale $H$ interacts with solid molecules of both membrane and substrate, which is similar to the description of a fluid confined within nanoporous materials (pores sizes less than 2 nm). Unlike bulk, the confined fluids exhibit inhomogeneous density distribution corresponding to layer adsorption on the solid surfaces. Also, the confined fluid pressure is significantly different from the bulk equation of state due to the influence of the well known disjoining pressure \cite{israelachvili2015intermolecular}. One of the most powerful approaches which can describe the thermodynamics and the packing properties of the confined fluid is c-DFT. In our work, we use a popular version of c-DFT, which was successfully applied to describe various nanoscale problems \cite{wu2006density}. Assuming most of the fluid molecules are stored in the flat part of the GNB ($r<L-2\delta$ in Figure~\ref{fig:profile_displacement}), the inner GNB fluid density $\rho(z)$ depends on the normal coordinate $z$ only. Thus, in accordance with c-DFT, the confined fluid energy (the grand potential) accounting for the interaction with both the membrane and the substrate can be written as:
\begin{align}
\label{eq:omega}
&\Omega[\rho(z)]=F_\text{id}[\rho(z)]+F_\text{attr}[\rho(z)]+F_\text{rep}[\rho(z)]+\nonumber 
\\  
&\quad
+\pi L^2\int_0^H dz \rho(z)\left[U(z)+U(H-z)-\mu\right]
\end{align}
where $\mu$ is the chemical potential defining the confined density; $U(z)$ and $U(H-z)$ the external potentials of the substrate and the membrane at the distance $H$; $F_\text{id}$, $F_\text{attr}$, $F_\text{rep}$ are the ideal gas contribution, the influences of molecular attraction and hard spheres repulsion, respectively. The detailed description of these terms and expressions for external potential can be found in the supplementary material of the work \cite{aslyamov2020model}. The confined fluid density distribution is defined by the minimization condition of the potential (\ref{eq:omega}):
\begin{equation}
\label{eq:density_distribution}
\frac{\delta \Omega[\rho(z)]}{\delta \rho(z)}=0
\end{equation}
In the studies of the bulk GNBs and the membranes \cite{yue2012analytical,khestanova2016universal,lyublinskaya2019effect, zhilyaev2019liquid} the confined fluid energy is equal to $-PV$, but it is true in the thermodynamic limit only $H/d>>1$ (where $d$ is the molecular diameter), since $\Omega\to-PV$. We use the finite volume expression $E_{cf}=\Omega[\rho(z)]$ and calculate the confined fluid pressure as:
\begin{equation}
\label{eq:confined_pressure}
P=-\frac{\partial \Omega}{\partial V}=-\int_0^H dz \rho(z) \frac{\partial U(z)}{\partial z}
\end{equation}
where $V=2\pi L^2 H$ is the pancake volume, and in the last equality, we used the results for the flat-geometry \cite{henderson2005statistical}.

In addition to momentum conservation (\ref{eq:momentum_conservation}) the mechanical equilibrium demands the minimization of the total energy (\ref{eq:total_energy}) over $H$ with a fixed radius $L$ \cite{yue2012analytical}:
\begin{equation}
\label{eq:energy_minimization}
\frac{\partial E_\text{tot}}{\partial H}=0%\frac{\partial E_\text{tot}}{\partial L}=0
\end{equation}
At the fixed radius $L$ we use the pressure definition (\ref{eq:confined_pressure}), so the more detailed form of the derivative of $E_\text{tot}$ over $H$ (\ref{eq:energy_minimization}) can be written as:
\begin{equation}
\label{eq:energy_minimization_detailed}
\frac{\partial E_\text{el}}{\partial H}-\pi L^2 P=0
\end{equation}
The defined expressions for the elastic energy (\ref{eq:elstic_energy}), (\ref{eq:strain_tensor}) allows to calculate the analytical solutions of (\ref{eq:energy_minimization_detailed}) at an arbitrary $L$. Assuming $L/\delta>2$ which is a very natural property of the nanoscale pancakes, condition \eqref{eq:energy_minimization_detailed} results in the following compact form:
\begin{align}
%\label{eq:displacement_equilibrium}
%&u_0=-\frac{\nu}{4\pi}\left(\frac{H}{L}\right)^2, \\
\label{eq:height_equilibrium}
P=\frac{11 H^3 Y(1-3 k/4)}{32 \pi^{3/2} \delta ^3 L \left(1-\nu ^2\right)}
%&H^3=\frac{2\pi^{3/2}\delta^3 L P (1-\nu^2)}{Y(1-3 k/4)}
\end{align}

\noindent To qualitatively verify relation \eqref{eq:height_equilibrium} we consider the similar problem — the indentation of a thin disk by cylinder with known radii $R_\text{out}$ and $R_\text{in}$, respectively. The major published results consider the limit case when the disk radius is much grater than the cylinder disk \cite{chandler2020indentation}. In the case of our interest when these radii are comparable an analytical result exists for $\nu=1/3$ only \cite{vella2017indentation}. Considering the limit $R_\text{in}\to R_\text{out}$ in expression (19) from \cite{vella2017indentation} (in our denotations) provides the relation $P\sim H^3/L\delta^3$ in agreement with our result \eqref{eq:height_equilibrium}.

Obtained expression %(\ref{eq:displacement_equilibrium}) and
(\ref{eq:height_equilibrium}) defines elastic energy and the relation between $H$ and $L$, which correspond to the pressure induced by the trapped molecules inside the pancake.
Thus, the isomass conditions of equilibrium pancakes are the following: 
\begin{align}
\label{eq:master_condition}
\pi L^2 \int_0^H\rho(z)dz=\text{const}\nonumber \\ 
L=\frac{11 H^3 Y (1-3 k/4)}{32 \pi^{3/2} \delta^3 P (1-\nu^2)}
\end{align}

% \begin{equation}
% \label{eq:master_condition}
% \begin{cases}
% \pi L^2 \int_0^H\rho(z)dz=\text{const}\nonumber \\
% \\
% L=\dfrac{H^3 Y (1-3 k/4)}{2 \pi^{3/2} \delta^3 P (1-\nu^2)}
% \end{cases}
% \end{equation}

To describe the stable states of the pancakes
we consider the total energy as a function of the height $H$ assuming the mechanical condition $L(H)$ and the constant mass of the trapped molecules inside GNB \eqref{eq:master_condition}.

\section{results and discussion}
\begin{figure*}
    \centering
     \includegraphics[width=\textwidth]{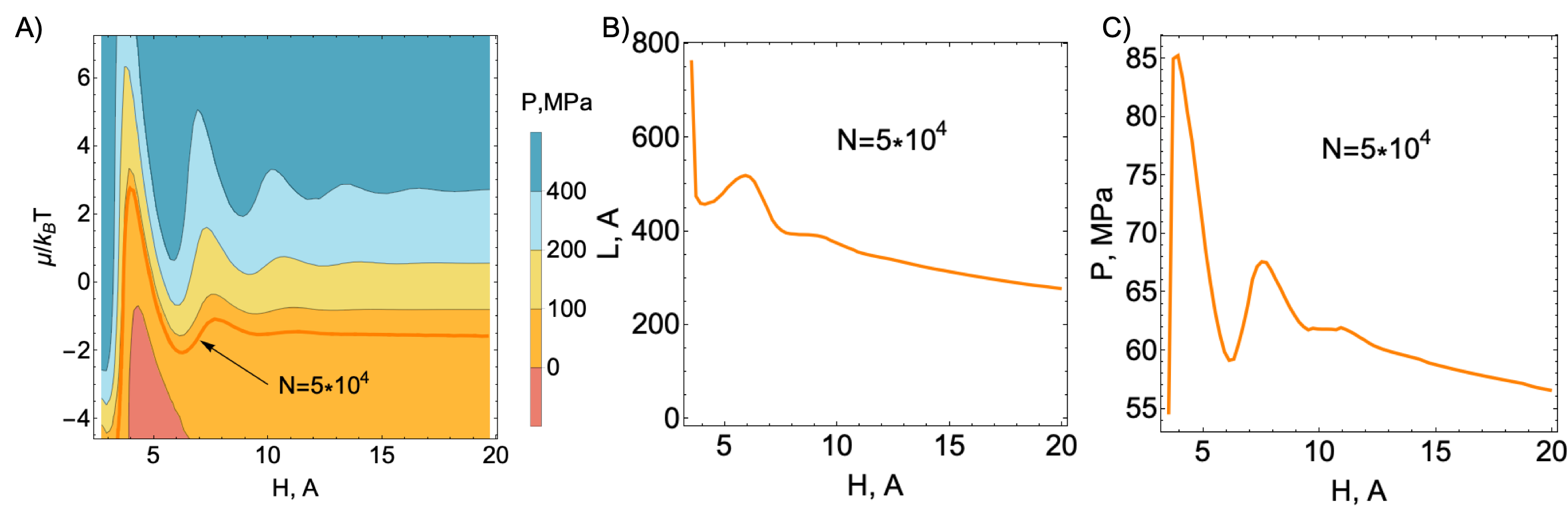}
    \caption{\label{fig:iso-mass}
    A) The contour plot of the confined pressure as function of the height and chemical potential calculated at $T=350K$. The orange solid line is the iso-mass curve, which is defined by the constant number of the trapped molecules ($50\times 10^4$) and the mechanical equilibrium conditions (\ref{eq:master_condition}). 
    B) The isomass curve from the left figure in terms of the height $H$ and the footprint radius $L$ exhibits the s-shape transition between one layer and two layers GNBs. 
    C) The confined fluid pressure exhibits oscillations and decreases as the GNB height becomes larger.}
\end{figure*}

We considered the GNBs, which consist of two graphene sheets and argon molecules inside at a wide temperature range $\SI{250}{K}\leq T\leq\SI{450}{K}$. Our results are dedicated to nanoscale GNB with the heights $H\leq 4d\sim \SI{1.2}{nm}$, where disjoining pressure plays the crucial role in the thermodynamics description \cite{gregoire2018estimation}. The trapped molecules inside pancakes exhibit the inhomogeneous distribution $\rho(z)$ and act on the graphene membrane with the pressure $P$, which can be defined by the fluid chemical potential and the height. The example of the confined pressure (\ref{eq:confined_pressure}) as a function of the height $H$ and the chemical potential $\mu$ is shown in Figure~\ref{fig:iso-mass}A illustrating the case of $n=5\times10^4$ argon molecules at $T=\SI{350}{K}$. The area of considered parameters contains the negative pressure region, which does not fit equilibrium pancakes condition with the positive $H$, $L$, and can be considered as a forbidden zone. Thus the isomass equilibrium conditions \eqref{eq:master_condition} define the locus of the pancakes parameters on the plane of $\mu$-$H$ as shown by the solid orange line in Figure~\ref{fig:iso-mass}A. 

\begin{figure*}
    \centering
    \includegraphics[width=\textwidth]{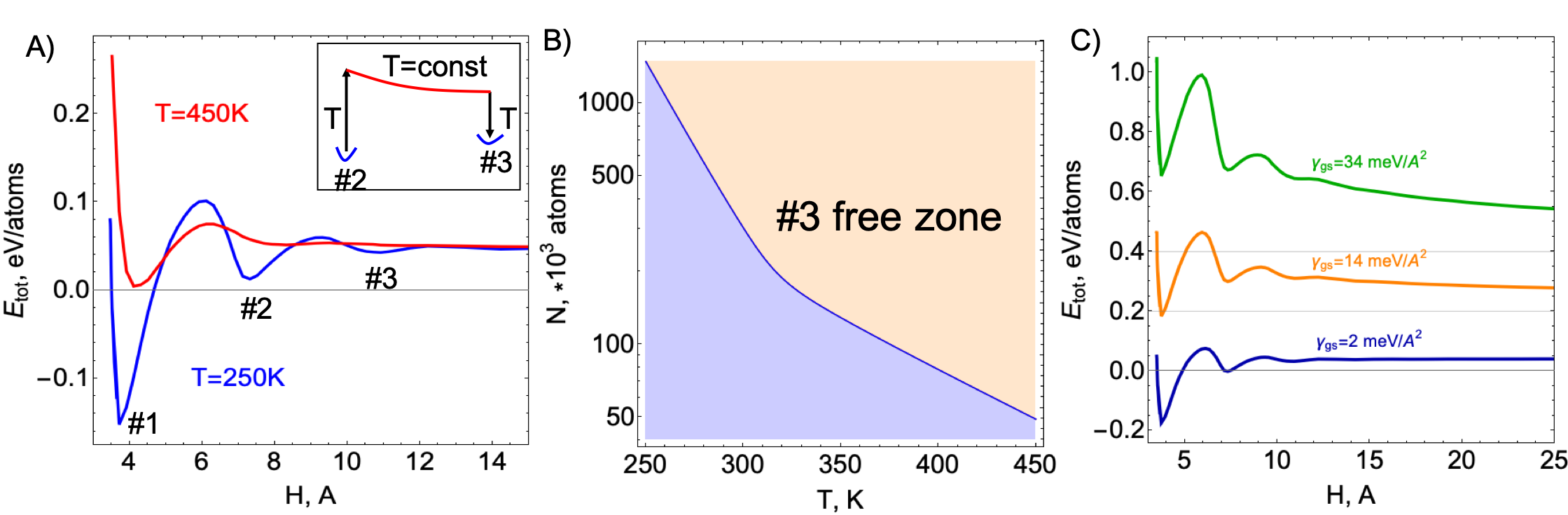}
    \caption{\label{fig:total_energy}
    A) The total energy of the pancake which contain $5\times10^4$ of argon molecules at various temperatures: $\SI{250}{K}$ (blue curve) and $\SI{450}{K}$ (red curve). The total energy minimums are denoted as \#1, \#2 and \#3 correspond to one-,two- and three layers GNB, respectively. Inset shows the scheme of temperature induces the transition from \#2 to \#3. 
    B) The blue zone corresponds to the relation $N<N_c(T)$, where three minimum points can be found. In the orange region \#3 minimum has already disappeared, and there is the three-layers GNB free zone. 
    C) The dependence of the total energy on the adhesion for the GNBs with 50 thousand argon molecules at \SI{250}{K}. Blue, orange and green curves correspond to various $\gamma_\text{gs}$ coefficients \SI{2}{meV/\AA^2}, \SI{14}{meV/\AA^2} and \SI{34}{meV/\AA^2}, respectively. The orange and green curves are shifted up by $\SI{0.2}{eV/atoms}$ and $\SI{0.4}{eV/atoms}$, respectively}
\end{figure*}

The iso-mass curve can be rewritten using the footprint radius $L$ as a function of the height $H$ (see Figure~\ref{fig:iso-mass}B). In terms of $L$ and $H$ the iso-mass curve shows the regions of the significant decrease of the footprint. At the same time, the height remains almost constant with the values, which are multiple of molecular diameter. Also, the transitions between the various values of $H$ exhibit S-shape initially and then become monotonically increasing function. Such transitions between the slopping $H$ regions may correspond to the transformation of one-molecular layer pancake to multilayer form with a decrease of the footprint radius. The corresponding to this system equilibrium pressure is shown in Figure~\ref{fig:iso-mass}C. Calculated from \eqref{eq:confined_pressure} pressure exhibits oscillation behavior and decreases as the height of the GNB becomes larger. Thus, the formation of the more pancake GNBs corresponds to the larger pressure. 

\begin{figure*}[t]
    \centering
    \includegraphics[width=\textwidth]{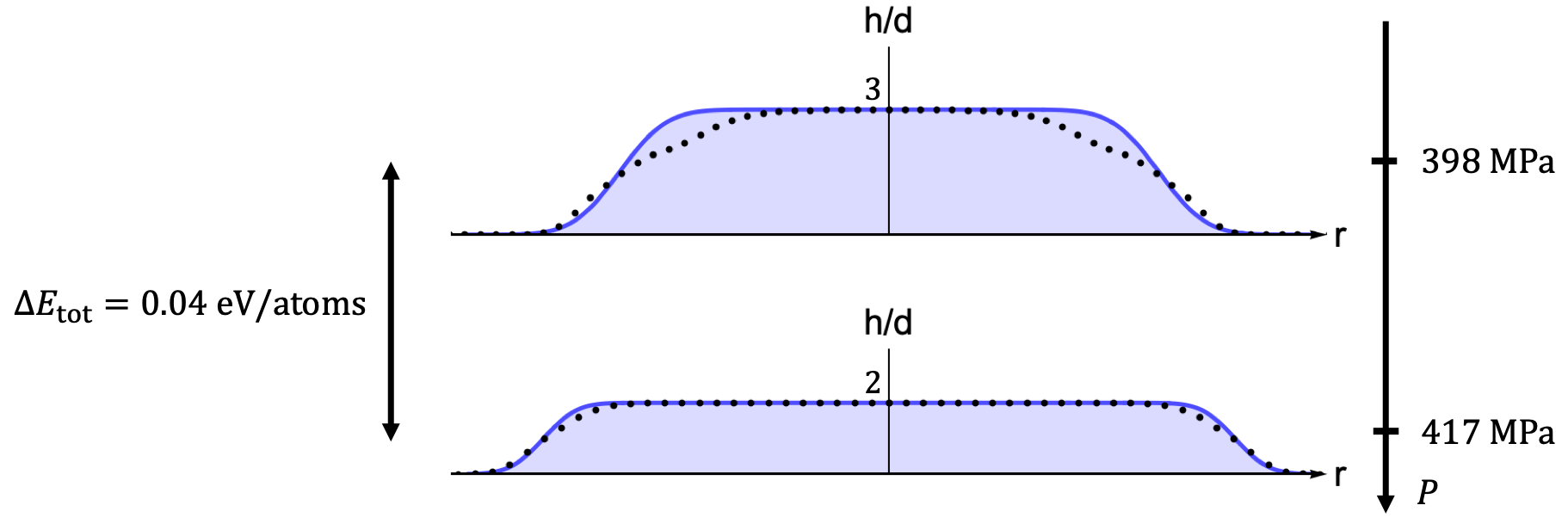}
    \caption{\label{fig:MDvsTheory}
    The comparison of the profiles (blue curves) and the thermodynamic parameters are calculated by proposed approach with the results (black dots) of previously published MD simulations. Both methods consider 2500 argon molecules between two graphene sheets with $\gamma_\text{gs}=\SI{17}{meV/\AA^2}$ at $T=\SI{300}{K}$. The profile parameters $k=1.187$ and $\delta_0= \SI{2.925}{\AA}$ provides exact fit of the corresponding MD pressures — \SI{417}{MPa} and \SI{398}{MPa} in the cases of two- and three-layers GNBs, respectively. The two-layers GNB is more stable, that also is in the agreement with the simulations. 
    } 
\end{figure*}

The iso-mass contour calculated from expressions (\ref{eq:master_condition}) defines the corresponding pressures (\ref{eq:confined_pressure}) and the density distributions (\ref{eq:density_distribution}). This information is enough to obtain all contributions to the total energy \eqref{eq:total_energy} as functions of $H$. Figure~\ref{fig:total_energy}A shows the total energy of the nanoscale pancakes containing $5\times10^4$ of argon molecules at various temperatures. As one can see from Figure~\ref{fig:total_energy}A the calculated curves become smother as the temperature increases. The coldest case $T=\SI{250}{K}$ exhibits three minima of the total energy divided by notable gaps, which significantly decrease or even disappear as the temperature becomes large. The calculated heights $H$ and density distributions demonstrate that \#1, \#2 and \#3 minima correspond to the pancakes containing one, two and three molecular layers, respectively. Thus, the total energy local minimums define metastable pancakes with the height is multiple of the diameter of the trapped molecule. 

We also investigate what other parameters besides the temperature have crucial influence on the GNB equilibrium states. As in work \cite{sanchez2018mechanics}, we consider the number of trapped molecules $N$ and the specific adhesion energy $\gamma_\text{gs}$. Figure~\ref{fig:total_energy}B demonstrates that for the amount of molecules $N$ larger certain $N_c(T)$ there is not \#3 minimum. Since Figure~\ref{fig:total_energy}B shows three-layers GNB free zone. More stable mono- and bi-layers, these formations disappear either as $N$ further increases. The adhesion energy $\gamma_\text{gs}$ influences similarly, that is shown in Figure~\ref{fig:total_energy}C. 
In the case of sufficiently small parameter $\gamma_\text{gs}\sim\SI{0.014}{eV/\A^2}$ the pancake configuration becomes more stable than the states with $H>3d$, then the trapped fluid is a continuum and the standard membrane theory can be used.
These observed properties qualitatively coincide with the analytical and simulation results of the work \cite{sanchez2018mechanics}, where authors describe the breakdown of the continuum model as the number of molecules or the adhesion energy decrease. Also, our calculations reveal the size of the vertical limit of the layered structure $H<H_c=3d$ due to the short-range effect of the disjoining pressure. This numerical estimation coincides well with the corresponding parameter observed for water-filled GNBs \cite{sanchez2018mechanics}. Thus, considering small trapped molecules, the maximal vertical size of the graphene pancakes is equal to three molecular diameters.

To verify the proposed profile geometry numerically, we used the results of the computer simulations of argon-filled GNB at $T=\SI{300}{K}$ and $\gamma_\text{gs}=\SI{17}{meV/\A^2}$. In the work \cite{iakovlev2017atomistic} was demonstrated that  2500 argon molecules between two graphene layers at $T=\SI{300}{K}$ might form more than one GNB state. More precisely, two and three layers GNBs with flat profiles were observed at pressures \SI{417}{MPa} and \SI{398}{MPa}, respectively.
We compared GNB profiles provided our theory with the results of the MD simulations. This comparison allows us to define the explicit form of the curved region to achieve the equilibrium pressures provided by MD. Figure~\ref{fig:MDvsTheory} shows the profiles corresponding to \#2 and \#3 total energy minima at a constant number of argon molecules $N=2500$. As one can see from Figure~\ref{fig:MDvsTheory} the values of the confined fluid pressures calculated for two and three layers coincide exactly with the corresponding MD data. The calculated profiles (blue line in Figure~\ref{fig:MDvsTheory}) fit the simulation results (black dots) well, especially in the case of GNB \#2. The theoretical profile of GNB\#3 shows the notable deviation from the MD result in the upper corners. However, our theory correctly provides the footprint radius, which crucially influences equilibrium via adhesion energy. Also, the sign of the energy difference between these states coincides with MD simulations, which means the two-layers GNB is more stable. %The illustration of the total energy functions in Figure~\ref{fig:total_energy}A shows the lack of global minimum in the range $H<3d$. This property results in the metastability of the pancake states \#1, \#2 and \#3 with the discrete vertical sizes. 

Our calculations are in agreement with published experiment %published by Ekaterina Khestanova in Ph.D. thesis 
\cite{khestanova2018van}. The small GNB ($N\sim 10000$) with pancake geometry is heated to various temperatures, while the profiles AFM measurements are carried out at fixed initial temperature after the cooling. The experiment shows the temperature-induced transitions between GNB configuration. More precisely, the height discretely increases (the step equals roughly to molecular diameter), and the footprint radius decreases as the temperature of previous heating becomes larger. The proposed metastable behavior can explain this result. Indeed the heating to sufficiently large temperature ruins the certain metastable pancake state, and after the cooling procedure, the system may return to another state with larger energy. In terms of Figure~\ref{fig:total_energy}A the heating to \SI{450}{K} induces the transition from \#2 at \SI{350}{K} to the direction of the energy decreasing as shown in the inset of Figure~\ref{fig:total_energy}A. After the cooling to \SI{350}{K} the GNB may obtain a new metastable state \#3 as the first on the way back. Thus, the measurements of the profiles at \SI{350}{K} before and after heating/cooling shows the transition from \#2 to \#3.

\section{conclusions}
We developed a theory of nanoscale pancakes of molecules trapped between graphene sheets accounting for confined fluid properties. The major differences of the pancakes from other existing geometries are discrete values of the vertical size with the step comparable to the molecular diameter and the existence at the subnanometer vertical scale ($H\leq3d$) only. Our theory reveals the mechanical equilibrium of such pancake profiles as a relation between the elastics stresses and confined fluids pressure. We demonstrated that the disjoining pressure results in the discrete size behavior and the lack of large flat profiles. Analysis of the total energy as a function of the pancake height shows the discrete set of local minima points corresponding to pancake profiles with well-structured layers of molecules inside. Our model qualitatively coincides with the published molecular dynamics simulations, which observed crucial dependence of the pancakes on the number of the trapped molecules and the value of adhesion energy. Also, the calculated profiles and pressures fit corresponding MD results well.

\begin{acknowledgments}
Timur Aslyamov is grateful to Aleksey Khlyupin for discussions and useful comments.
\end{acknowledgments}

\nocite{*}

\bibliography{sample}% Produces the bibliography via BibTeX.

\begin{thebibliography}{19}%
\makeatletter
\providecommand \@ifxundefined [1]{%
 \@ifx{#1\undefined}
}%
\providecommand \@ifnum [1]{%
 \ifnum #1\expandafter \@firstoftwo
 \else \expandafter \@secondoftwo
 \fi
}%
\providecommand \@ifx [1]{%
 \ifx #1\expandafter \@firstoftwo
 \else \expandafter \@secondoftwo
 \fi
}%
\providecommand \natexlab [1]{#1}%
\providecommand \enquote  [1]{``#1''}%
\providecommand \bibnamefont  [1]{#1}%
\providecommand \bibfnamefont [1]{#1}%
\providecommand \citenamefont [1]{#1}%
\providecommand \href@noop [0]{\@secondoftwo}%
\providecommand \href [0]{\begingroup \@sanitize@url \@href}%
\providecommand \@href[1]{\@@startlink{#1}\@@href}%
\providecommand \@@href[1]{\endgroup#1\@@endlink}%
\providecommand \@sanitize@url [0]{\catcode `\\12\catcode `\$12\catcode
  `\&12\catcode `\#12\catcode `\^12\catcode `\_12\catcode `\%12\relax}%
\providecommand \@@startlink[1]{}%
\providecommand \@@endlink[0]{}%
\providecommand \url  [0]{\begingroup\@sanitize@url \@url }%
\providecommand \@url [1]{\endgroup\@href {#1}{\urlprefix }}%
\providecommand \urlprefix  [0]{URL }%
\providecommand \Eprint [0]{\href }%
\providecommand \doibase [0]{http://dx.doi.org/}%
\providecommand \selectlanguage [0]{\@gobble}%
\providecommand \bibinfo  [0]{\@secondoftwo}%
\providecommand \bibfield  [0]{\@secondoftwo}%
\providecommand \translation [1]{[#1]}%
\providecommand \BibitemOpen [0]{}%
\providecommand \bibitemStop [0]{}%
\providecommand \bibitemNoStop [0]{.\EOS\space}%
\providecommand \EOS [0]{\spacefactor3000\relax}%
\providecommand \BibitemShut  [1]{\csname bibitem#1\endcsname}%
\let\auto@bib@innerbib\@empty
%</preamble>
\bibitem [{\citenamefont {Khestanova}\ \emph {et~al.}(2016)\citenamefont
  {Khestanova}, \citenamefont {Guinea}, \citenamefont {Fumagalli},
  \citenamefont {Geim},\ and\ \citenamefont
  {Grigorieva}}]{khestanova2016universal}%
  \BibitemOpen
  \bibfield  {author} {\bibinfo {author} {\bibfnamefont {E.}~\bibnamefont
  {Khestanova}}, \bibinfo {author} {\bibfnamefont {F.}~\bibnamefont {Guinea}},
  \bibinfo {author} {\bibfnamefont {L.}~\bibnamefont {Fumagalli}}, \bibinfo
  {author} {\bibfnamefont {A.}~\bibnamefont {Geim}}, \ and\ \bibinfo {author}
  {\bibfnamefont {I.}~\bibnamefont {Grigorieva}},\ }\href@noop {} {\bibfield
  {journal} {\bibinfo  {journal} {Nature communications}\ }\textbf {\bibinfo
  {volume} {7}},\ \bibinfo {pages} {12587} (\bibinfo {year}
  {2016})}\BibitemShut {NoStop}%
\bibitem [{\citenamefont {Dai}\ \emph {et~al.}(2018)\citenamefont {Dai},
  \citenamefont {Hou}, \citenamefont {Sanchez}, \citenamefont {Wang},
  \citenamefont {Brennan}, \citenamefont {Zhang}, \citenamefont {Liu},\ and\
  \citenamefont {Lu}}]{dai2018interface}%
  \BibitemOpen
  \bibfield  {author} {\bibinfo {author} {\bibfnamefont {Z.}~\bibnamefont
  {Dai}}, \bibinfo {author} {\bibfnamefont {Y.}~\bibnamefont {Hou}}, \bibinfo
  {author} {\bibfnamefont {D.~A.}\ \bibnamefont {Sanchez}}, \bibinfo {author}
  {\bibfnamefont {G.}~\bibnamefont {Wang}}, \bibinfo {author} {\bibfnamefont
  {C.~J.}\ \bibnamefont {Brennan}}, \bibinfo {author} {\bibfnamefont
  {Z.}~\bibnamefont {Zhang}}, \bibinfo {author} {\bibfnamefont
  {L.}~\bibnamefont {Liu}}, \ and\ \bibinfo {author} {\bibfnamefont
  {N.}~\bibnamefont {Lu}},\ }\href@noop {} {\bibfield  {journal} {\bibinfo
  {journal} {Physical review letters}\ }\textbf {\bibinfo {volume} {121}},\
  \bibinfo {pages} {266101} (\bibinfo {year} {2018})}\BibitemShut {NoStop}%
\bibitem [{\citenamefont {Yue}\ \emph {et~al.}(2012)\citenamefont {Yue},
  \citenamefont {Gao}, \citenamefont {Huang},\ and\ \citenamefont
  {Liechti}}]{yue2012analytical}%
  \BibitemOpen
  \bibfield  {author} {\bibinfo {author} {\bibfnamefont {K.}~\bibnamefont
  {Yue}}, \bibinfo {author} {\bibfnamefont {W.}~\bibnamefont {Gao}}, \bibinfo
  {author} {\bibfnamefont {R.}~\bibnamefont {Huang}}, \ and\ \bibinfo {author}
  {\bibfnamefont {K.~M.}\ \bibnamefont {Liechti}},\ }\href@noop {} {\bibfield
  {journal} {\bibinfo  {journal} {Journal of Applied Physics}\ }\textbf
  {\bibinfo {volume} {112}},\ \bibinfo {pages} {083512} (\bibinfo {year}
  {2012})}\BibitemShut {NoStop}%
\bibitem [{\citenamefont {Iakovlev}\ \emph {et~al.}(2017)\citenamefont
  {Iakovlev}, \citenamefont {Zhilyaev},\ and\ \citenamefont
  {Akhatov}}]{iakovlev2017atomistic}%
  \BibitemOpen
  \bibfield  {author} {\bibinfo {author} {\bibfnamefont {E.}~\bibnamefont
  {Iakovlev}}, \bibinfo {author} {\bibfnamefont {P.}~\bibnamefont {Zhilyaev}},
  \ and\ \bibinfo {author} {\bibfnamefont {I.}~\bibnamefont {Akhatov}},\
  }\href@noop {} {\bibfield  {journal} {\bibinfo  {journal} {Scientific
  reports}\ }\textbf {\bibinfo {volume} {7}},\ \bibinfo {pages} {17906}
  (\bibinfo {year} {2017})}\BibitemShut {NoStop}%
\bibitem [{\citenamefont {Khestanova}(2018)}]{khestanova2018van}%
  \BibitemOpen
  \bibfield  {author} {\bibinfo {author} {\bibfnamefont {E.}~\bibnamefont
  {Khestanova}},\ }\emph {\bibinfo {title} {Van der Waals heterostructures:
  fabrication, mechanical and electronic properties}},\ \href@noop {} {Ph.D.
  thesis},\ \bibinfo  {school} {The University of Manchester (United Kingdom)}
  (\bibinfo {year} {2018})\BibitemShut {NoStop}%
\bibitem [{\citenamefont {Sanchez}\ \emph {et~al.}(2018)\citenamefont
  {Sanchez}, \citenamefont {Dai}, \citenamefont {Wang}, \citenamefont
  {Cantu-Chavez}, \citenamefont {Brennan}, \citenamefont {Huang},\ and\
  \citenamefont {Lu}}]{sanchez2018mechanics}%
  \BibitemOpen
  \bibfield  {author} {\bibinfo {author} {\bibfnamefont {D.~A.}\ \bibnamefont
  {Sanchez}}, \bibinfo {author} {\bibfnamefont {Z.}~\bibnamefont {Dai}},
  \bibinfo {author} {\bibfnamefont {P.}~\bibnamefont {Wang}}, \bibinfo {author}
  {\bibfnamefont {A.}~\bibnamefont {Cantu-Chavez}}, \bibinfo {author}
  {\bibfnamefont {C.~J.}\ \bibnamefont {Brennan}}, \bibinfo {author}
  {\bibfnamefont {R.}~\bibnamefont {Huang}}, \ and\ \bibinfo {author}
  {\bibfnamefont {N.}~\bibnamefont {Lu}},\ }\href@noop {} {\bibfield  {journal}
  {\bibinfo  {journal} {Proceedings of the National Academy of Sciences}\
  }\textbf {\bibinfo {volume} {115}},\ \bibinfo {pages} {7884} (\bibinfo {year}
  {2018})}\BibitemShut {NoStop}%
\bibitem [{\citenamefont
  {Israelachvili}(2015)}]{israelachvili2015intermolecular}%
  \BibitemOpen
  \bibfield  {author} {\bibinfo {author} {\bibfnamefont {J.~N.}\ \bibnamefont
  {Israelachvili}},\ }\href@noop {} {\emph {\bibinfo {title} {Intermolecular
  and surface forces}}}\ (\bibinfo  {publisher} {Academic press},\ \bibinfo
  {year} {2015})\BibitemShut {NoStop}%
\bibitem [{\citenamefont {Long}\ \emph {et~al.}(2011)\citenamefont {Long},
  \citenamefont {Palmer}, \citenamefont {Coasne}, \citenamefont
  {{\'S}liwinska-Bartkowiak},\ and\ \citenamefont
  {Gubbins}}]{long2011pressure}%
  \BibitemOpen
  \bibfield  {author} {\bibinfo {author} {\bibfnamefont {Y.}~\bibnamefont
  {Long}}, \bibinfo {author} {\bibfnamefont {J.~C.}\ \bibnamefont {Palmer}},
  \bibinfo {author} {\bibfnamefont {B.}~\bibnamefont {Coasne}}, \bibinfo
  {author} {\bibfnamefont {M.}~\bibnamefont {{\'S}liwinska-Bartkowiak}}, \ and\
  \bibinfo {author} {\bibfnamefont {K.~E.}\ \bibnamefont {Gubbins}},\
  }\href@noop {} {\bibfield  {journal} {\bibinfo  {journal} {Physical Chemistry
  Chemical Physics}\ }\textbf {\bibinfo {volume} {13}},\ \bibinfo {pages}
  {17163} (\bibinfo {year} {2011})}\BibitemShut {NoStop}%
\bibitem [{\citenamefont {Svetovoy}\ \emph {et~al.}(2016)\citenamefont
  {Svetovoy}, \citenamefont {Devic}, \citenamefont {Snoeijer},\ and\
  \citenamefont {Lohse}}]{svetovoy2016effect}%
  \BibitemOpen
  \bibfield  {author} {\bibinfo {author} {\bibfnamefont {V.~B.}\ \bibnamefont
  {Svetovoy}}, \bibinfo {author} {\bibfnamefont {I.}~\bibnamefont {Devic}},
  \bibinfo {author} {\bibfnamefont {J.~H.}\ \bibnamefont {Snoeijer}}, \ and\
  \bibinfo {author} {\bibfnamefont {D.}~\bibnamefont {Lohse}},\ }\href@noop {}
  {\bibfield  {journal} {\bibinfo  {journal} {Langmuir}\ }\textbf {\bibinfo
  {volume} {32}},\ \bibinfo {pages} {11188} (\bibinfo {year}
  {2016})}\BibitemShut {NoStop}%
\bibitem [{\citenamefont {Kim}\ \emph {et~al.}(1999)\citenamefont {Kim},
  \citenamefont {Mate}, \citenamefont {Hannibal},\ and\ \citenamefont
  {Perry}}]{kim1999disjoining}%
  \BibitemOpen
  \bibfield  {author} {\bibinfo {author} {\bibfnamefont {H.~I.}\ \bibnamefont
  {Kim}}, \bibinfo {author} {\bibfnamefont {C.~M.}\ \bibnamefont {Mate}},
  \bibinfo {author} {\bibfnamefont {K.~A.}\ \bibnamefont {Hannibal}}, \ and\
  \bibinfo {author} {\bibfnamefont {S.~S.}\ \bibnamefont {Perry}},\ }\href@noop
  {} {\bibfield  {journal} {\bibinfo  {journal} {Physical review letters}\
  }\textbf {\bibinfo {volume} {82}},\ \bibinfo {pages} {3496} (\bibinfo {year}
  {1999})}\BibitemShut {NoStop}%
\bibitem [{\citenamefont {Gr{\'e}goire}\ \emph {et~al.}(2018)\citenamefont
  {Gr{\'e}goire}, \citenamefont {Malheiro},\ and\ \citenamefont
  {Miqueu}}]{gregoire2018estimation}%
  \BibitemOpen
  \bibfield  {author} {\bibinfo {author} {\bibfnamefont {D.}~\bibnamefont
  {Gr{\'e}goire}}, \bibinfo {author} {\bibfnamefont {C.}~\bibnamefont
  {Malheiro}}, \ and\ \bibinfo {author} {\bibfnamefont {C.}~\bibnamefont
  {Miqueu}},\ }\href@noop {} {\bibfield  {journal} {\bibinfo  {journal}
  {Continuum Mechanics and Thermodynamics}\ }\textbf {\bibinfo {volume} {30}},\
  \bibinfo {pages} {347} (\bibinfo {year} {2018})}\BibitemShut {NoStop}%
\bibitem [{\citenamefont {Wu}(2006)}]{wu2006density}%
  \BibitemOpen
  \bibfield  {author} {\bibinfo {author} {\bibfnamefont {J.}~\bibnamefont
  {Wu}},\ }\href@noop {} {\bibfield  {journal} {\bibinfo  {journal} {AIChE
  journal}\ }\textbf {\bibinfo {volume} {52}},\ \bibinfo {pages} {1169}
  (\bibinfo {year} {2006})}\BibitemShut {NoStop}%
\bibitem [{\citenamefont {Aslyamov}\ \emph {et~al.}(2020)\citenamefont
  {Aslyamov}, \citenamefont {Iakovlev}, \citenamefont {Akhatov},\ and\
  \citenamefont {Zhilyaev}}]{aslyamov2020model}%
  \BibitemOpen
  \bibfield  {author} {\bibinfo {author} {\bibfnamefont {T.}~\bibnamefont
  {Aslyamov}}, \bibinfo {author} {\bibfnamefont {E.}~\bibnamefont {Iakovlev}},
  \bibinfo {author} {\bibfnamefont {I.~S.}\ \bibnamefont {Akhatov}}, \ and\
  \bibinfo {author} {\bibfnamefont {P.}~\bibnamefont {Zhilyaev}},\ }\href@noop
  {} {\bibfield  {journal} {\bibinfo  {journal} {The Journal of Chemical
  Physics}\ }\textbf {\bibinfo {volume} {152}},\ \bibinfo {pages} {054705}
  (\bibinfo {year} {2020})}\BibitemShut {NoStop}%
\bibitem [{\citenamefont {Wang}\ \emph {et~al.}(2013)\citenamefont {Wang},
  \citenamefont {Gao}, \citenamefont {Cao}, \citenamefont {Liechti},\ and\
  \citenamefont {Huang}}]{wang2013numerical}%
  \BibitemOpen
  \bibfield  {author} {\bibinfo {author} {\bibfnamefont {P.}~\bibnamefont
  {Wang}}, \bibinfo {author} {\bibfnamefont {W.}~\bibnamefont {Gao}}, \bibinfo
  {author} {\bibfnamefont {Z.}~\bibnamefont {Cao}}, \bibinfo {author}
  {\bibfnamefont {K.~M.}\ \bibnamefont {Liechti}}, \ and\ \bibinfo {author}
  {\bibfnamefont {R.}~\bibnamefont {Huang}},\ }\href@noop {} {\bibfield
  {journal} {\bibinfo  {journal} {Journal of Applied Mechanics}\ }\textbf
  {\bibinfo {volume} {80}},\ \bibinfo {pages} {040905} (\bibinfo {year}
  {2013})}\BibitemShut {NoStop}%
\bibitem [{\citenamefont {Lyublinskaya}\ \emph {et~al.}(2019)\citenamefont
  {Lyublinskaya}, \citenamefont {Babkin},\ and\ \citenamefont
  {Burmistrov}}]{lyublinskaya2019effect}%
  \BibitemOpen
  \bibfield  {author} {\bibinfo {author} {\bibfnamefont {A.}~\bibnamefont
  {Lyublinskaya}}, \bibinfo {author} {\bibfnamefont {S.}~\bibnamefont
  {Babkin}}, \ and\ \bibinfo {author} {\bibfnamefont {I.}~\bibnamefont
  {Burmistrov}},\ }\href@noop {} {\bibfield  {journal} {\bibinfo  {journal}
  {arXiv preprint arXiv:1909.12650}\ } (\bibinfo {year} {2019})}\BibitemShut
  {NoStop}%
\bibitem [{\citenamefont {Zhilyaev}\ \emph {et~al.}(2019)\citenamefont
  {Zhilyaev}, \citenamefont {Iakovlev},\ and\ \citenamefont
  {Akhatov}}]{zhilyaev2019liquid}%
  \BibitemOpen
  \bibfield  {author} {\bibinfo {author} {\bibfnamefont {P.}~\bibnamefont
  {Zhilyaev}}, \bibinfo {author} {\bibfnamefont {E.}~\bibnamefont {Iakovlev}},
  \ and\ \bibinfo {author} {\bibfnamefont {I.}~\bibnamefont {Akhatov}},\
  }\href@noop {} {\bibfield  {journal} {\bibinfo  {journal} {Nanotechnology}\
  }\textbf {\bibinfo {volume} {30}},\ \bibinfo {pages} {215701} (\bibinfo
  {year} {2019})}\BibitemShut {NoStop}%
\bibitem [{\citenamefont {Henderson}(2005)}]{henderson2005statistical}%
  \BibitemOpen
  \bibfield  {author} {\bibinfo {author} {\bibfnamefont {J.}~\bibnamefont
  {Henderson}},\ }\href@noop {} {\bibfield  {journal} {\bibinfo  {journal}
  {Physical Review E}\ }\textbf {\bibinfo {volume} {72}},\ \bibinfo {pages}
  {051602} (\bibinfo {year} {2005})}\BibitemShut {NoStop}%
\bibitem [{\citenamefont {Chandler}\ and\ \citenamefont
  {Vella}(2020)}]{chandler2020indentation}%
  \BibitemOpen
  \bibfield  {author} {\bibinfo {author} {\bibfnamefont {T.~G.}\ \bibnamefont
  {Chandler}}\ and\ \bibinfo {author} {\bibfnamefont {D.}~\bibnamefont
  {Vella}},\ }\href@noop {} {\bibfield  {journal} {\bibinfo  {journal} {arXiv
  preprint arXiv:2002.05634}\ } (\bibinfo {year} {2020})}\BibitemShut {NoStop}%
\bibitem [{\citenamefont {Vella}\ and\ \citenamefont
  {Davidovitch}(2017)}]{vella2017indentation}%
  \BibitemOpen
  \bibfield  {author} {\bibinfo {author} {\bibfnamefont {D.}~\bibnamefont
  {Vella}}\ and\ \bibinfo {author} {\bibfnamefont {B.}~\bibnamefont
  {Davidovitch}},\ }\href@noop {} {\bibfield  {journal} {\bibinfo  {journal}
  {Soft Matter}\ }\textbf {\bibinfo {volume} {13}},\ \bibinfo {pages} {2264}
  (\bibinfo {year} {2017})}\BibitemShut {NoStop}%
\end{thebibliography}%

\end{document}